# 量子密碼學中間人攻擊

Abel C. H. Chen
*Information & Communications Security Laboratory,*
*Chunghwa Telecom Laboratories*
Taoyuan, Taiwan
ORCID: 0000-0003-3628-3033

*摘要*—隨著 Google 研發出 Willow 量子晶片，帶動起量子計算的一波熱潮。在量子計算技術日益成熟的環境，量子安全通訊也逐漸受到重視。為建立量子安全通訊，已有多個量子金鑰分配協定被提出，例如：BB84 協定等，通過量子疊加態等特性來建立安全傳輸。然而，現行的量子金鑰分配協定可能在一定情境下會面臨威脅，所以本研究主要提出兩種中間人攻擊技巧，並且通過實例來論證如何破解量子密碼學。除此之外，本研究亦討論如何防範這兩種中間人攻擊，以及提升量子密碼學安全性的方法，可作為後續部署量子安全通訊之參考。

*關鍵字—量子密碼學、量子金鑰分配、BB84 協定、中間人攻擊*

## I. 前言

近年來，在 Google [1]、IBM [2]、微軟[3]等公司致力於量子計算技術的研發上，量子計算的硬體和軟體技術已經逐漸成熟。各式各樣的量子計算應用也陸續被發展出來，包含有量子神經網路應用於雲端運算[4]、量子神經網路應用於腦機介面[5]、量子神經網路應用於無線資源最佳化[6]、以及聯邦量子神經網路應用於無線通訊[7]等。然而，隨著量子計算應用的普及，量子安全通訊也開始受到重視，如何建立安全的量子通訊環境，將會是重要的研究方向之一。

目前已經有多個量子隨機數產生器[8]、量子金鑰分配協定[9]被提出，其中知名的量子金鑰分配協定包含有 BB84 協定[10]等。專家們巧妙運用量子特性，包含量子疊加態、量子糾纏態等，讓量子訊號可以在量子通訊傳輸過程中即使被監聽，監聽者也無法解譯出正確的資訊，並且傳送端和接收端雙方可以知道有被監聽，達到安全的量子通訊。然而，陸續有研究對量子金鑰分配協定可能存在的漏洞或限制提出可能的攻擊技巧，包含有實體層攻擊(Physical-Layer Attack)[11]、旁通道攻擊(Side-Channel Attack)[12]、特洛依木馬攻擊(Trojan-Horse Attack)[13]等。

有鑑於此，本研究著重在量子金鑰分配協定的中間人攻擊，提出兩種可能攻擊技巧，並且通過實際案例來證實可以攻破現行的量子金鑰分配協定。此外，本研究也對提出的兩種攻擊技巧提出防禦的策略，來避免量子密碼學中間人攻擊。本研究貢獻條列如下：

1. 本研究提出當發送端多次發送量子訊號的情況下，可能發生的中間人攻擊，並且提供實際案例說明。

2. 本研究提出當接收端提早發送控制訊號的情況下，可能發生的中間人攻擊，並且提供實際案例說明。

3. 本研究對量子密碼學中間人攻擊提出防禦策略，避免量子金鑰分配協定通訊過程中的中間人攻擊。

本研究共分為五個章節，第 II 節將介紹量子計算和現行的量子金鑰分配協定。第 III 節將對現行的量子金鑰分配協定提出中間人攻擊技巧，以及第 IV 節提出防禦策略來防範中間人攻擊。最後，第 V 節總結本研究貢獻，並且提出未來的研究方向。

## II. 研究背景

本節將先介紹量子計算，定義量子態和描述會在量子金鑰分配協定使用到的量子邏輯閘。然後再介紹現行的量子金鑰分配協定方法及其安全性討論。

### A. 量子計算

*1) 量子態定義*

量子位元與經典位元最大的差異在於量子位元存在疊加態，有$|v_0|^2$的機率是 0 且有$|v_1|^2$的機率是 1；而經典位元則不是 0 就是 1，同時只會是 0 或 1。因此，在表述量子位元的量子態時通常會用向量矩陣$\begin{bmatrix}v_0\\v_1\end{bmatrix}$的方式來表示，如公式(1)所示；其中，$v_0$表示為 0 的向量，$v_1$表示為 1 的向量。例如，當$|q\rangle = |0\rangle$時，則向量矩陣是$\begin{bmatrix}1\\0\end{bmatrix}$，如公式(2)所示；當$|q\rangle = |1\rangle$時，則向量矩陣是$\begin{bmatrix}0\\1\end{bmatrix}$，如公式(3)所示[14]。

$$|q\rangle = v_0|0\rangle + v_1|1\rangle \to \begin{bmatrix}v_0\\v_1\end{bmatrix}. \quad (1)$$

$$|q\rangle = |0\rangle = 1|0\rangle + 0|1\rangle \to |q\rangle = \begin{bmatrix}1\\0\end{bmatrix}. \quad (2)$$

$$|q\rangle = |1\rangle = 0|0\rangle + 1|1\rangle \to |q\rangle = \begin{bmatrix}0\\1\end{bmatrix}. \quad (3)$$

*2) 量子邏輯閘*

本研究主要分析的量子金鑰分配協定是採用 BB84 協定，在 BB84 協定主要會用到的量子邏輯閘主要包含有 Pauli-X 閘和 Hadamard 閘，所以本節主要介紹這兩個量子邏輯閘。

Pauli-X 閘主要可以產生 NOT 的效果，表示式如公式(4)所示。當對量子位元$|q\rangle$操作 Pauli-X 閘後可以變換向量$v_0$和向量$v_1$，從$\begin{bmatrix}v_0\\v_1\end{bmatrix}$變換為$\begin{bmatrix}v_1\\v_0\end{bmatrix}$，如公式(5)所示[14]。由於每個量子位元初始值皆為$|0\rangle$，可以通過操作Pauli-X 閘來產生欲傳送的量子訊號。

$$X = \begin{bmatrix}0 & 1\\1 & 0\end{bmatrix}. \quad (4)$$

$$X|q\rangle = \begin{bmatrix}0 & 1\\1 & 0\end{bmatrix}\begin{bmatrix}v_0\\v_1\end{bmatrix} = \begin{bmatrix}v_1\\v_0\end{bmatrix}. \quad (5)$$

Hadamard 閘主要可以產生均勻疊加態的效果，表示式如公式(6)所示。並且，Hadamard閘是本身的逆，所以當操作兩次 Hadamard閘後可以還原為原本的值，如公式

(7)所示[14]。公式(8)表示對量子位元$|q\rangle$操作 Hadamard 閘後的結果；當$|q\rangle = |0\rangle$時，對量子位元$|q\rangle$操作 Hadamard 閘後是$\begin{bmatrix}1\\0\end{bmatrix}$，如公式(9)所示；當$|q\rangle = |1\rangle$時，對量子位元$|q\rangle$操作 Hadamard 閘後是$\begin{bmatrix}0\\1\end{bmatrix}$，如公式(10)所示。可以發現不論量子位元值為$|0\rangle$或$|1\rangle$時，操作 Hadamard 閘後都會變成 0 的機率是$\frac{1}{2}$且 1 的機率是$\frac{1}{2}$。由於不論量子位元值為$|0\rangle$或$|1\rangle$，操作 Hadamard 閘後 0 和 1 的機率會是各$\frac{1}{2}$，所以可以用來隱藏資訊，避免被監聽。

$$H = \frac{1}{\sqrt{2}}\begin{bmatrix}1 & 1\\1 & -1\end{bmatrix}. \tag{6}$$

$$HH = \frac{1}{\sqrt{2}}\begin{bmatrix}1 & 1\\1 & -1\end{bmatrix}\frac{1}{\sqrt{2}}\begin{bmatrix}1 & 1\\1 & -1\end{bmatrix},$$
$$= \frac{1}{2}\begin{bmatrix}2 & 0\\0 & 2\end{bmatrix} = \begin{bmatrix}1 & 0\\0 & 1\end{bmatrix}. \tag{7}$$

$$H|q\rangle = \frac{1}{\sqrt{2}}\begin{bmatrix}1 & 1\\1 & -1\end{bmatrix}\begin{bmatrix}v_0\\v_1\end{bmatrix} = \frac{1}{\sqrt{2}}\begin{bmatrix}v_0+v_1\\v_0-v_1\end{bmatrix}. \tag{8}$$

$$H|0\rangle = \frac{1}{\sqrt{2}}\begin{bmatrix}1 & 1\\1 & -1\end{bmatrix}\begin{bmatrix}1\\0\end{bmatrix} = \frac{1}{\sqrt{2}}\begin{bmatrix}1\\1\end{bmatrix},$$
$$= \begin{bmatrix}\frac{1}{\sqrt{2}}\\\frac{1}{\sqrt{2}}\end{bmatrix} = |U^+\rangle, \text{where } |U^+\rangle \in |U\rangle. \tag{9}$$

$$H|1\rangle = \frac{1}{\sqrt{2}}\begin{bmatrix}1 & 1\\1 & -1\end{bmatrix}\begin{bmatrix}0\\1\end{bmatrix} = \frac{1}{\sqrt{2}}\begin{bmatrix}1\\-1\end{bmatrix},$$
$$= \begin{bmatrix}\frac{1}{\sqrt{2}}\\\frac{-1}{\sqrt{2}}\end{bmatrix} = |U^-\rangle, \text{where } |U^-\rangle \in |U\rangle. \tag{10}$$

*B. 量子金鑰分配*

目前已經有多個量子金鑰分配協定被提出，包含運用量子疊加態的 BB84 協定[10],[15]-[16]和運用量子糾纏態的 E91 協定[17]-[19]等。由於本研究主要針對 BB84 協定設計中間人攻擊，所以本節聚焦於介紹 BB84 協定及其安全性討論。

*1) BB84 協定*

在BB84 協定基礎上，當 Alice 和 Bob 想協商$n$個位元的訊息(例如：$n$個位元的進階加密標準(Advanced Encryption Standard, AES)金鑰)，則需要用到$2\times n$個量子位元來傳輸。假設 Alice 為傳送端，而 Bob 為接收端，BB84 協定詳細流程如圖 1 所示，具體步驟描述如下。

Alice 首先依據下列步驟產生和傳送量子訊號：

(a). Alice 隨機產生$2\times n$個量子位元值，表示式為$Q_A = \{q_{A,2n-1}, q_{A,2n-2}, \ldots, q_{A,1}, q_{A,0}\}$；其中，每個量子位元初始值皆為$|0\rangle$，每個量子位元有$\frac{1}{2}$的機率會被操作 Pauli-X 閘變換為$|1\rangle$。

(b). Alice 隨機產生$2\times n$個控制訊號，表示式為$A = \{a_{2n-1}, a_{2n-2}, \ldots, a_1, a_0\}$；其中，第$i$個控制訊號$a_i$決定第$i$個量子位元$q_{A,i}$是否操作Hadamard閘，每個量子位元有$\frac{1}{2}$的機率會被操作 Hadamard 閘。

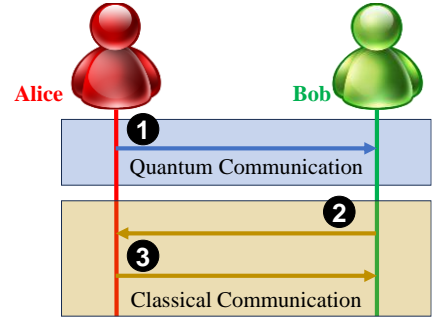

Fig. 1. BB84 協定

(c). Alice 對$2\times n$個量子位元值$Q_A$根據$2\times n$個控制訊號$A$各別操作後產生欲傳送的量子訊號$Q_T = \{q_{T,2n-1}, q_{T,2n-2}, \ldots, q_{T,1}, q_{T,0}\}$，並且通過量子通訊傳送量子訊號$Q_T$(即圖 1 中的 ❶ )給 Bob。

Bob 依據下列步驟量測量子訊號和傳送控制訊號：

(d). Bob 隨機產生$2\times n$個控制訊號，表示式為$B = \{b_{2n-1}, b_{2n-2}, \ldots, b_1, b_0\}$；其中，第$i$個控制訊號$b_i$決定第$i$個量子位元$q_{T,i}$是否操作 Hadamard 閘，每個量子位元有$\frac{1}{2}$的機率會被操作 Hadamard 閘。

(e). Bob 對$2\times n$個量子位元值$Q_T$根據$2\times n$個控制訊號$B$各別操作後產生量測量子訊號$Q_B = \{q_{B,2\times n-1}, q_{B,2\times n-2}, \ldots, q_{B,1}, q_{B,0}\}$，並且通過經典通訊傳送控制訊號$B$ (即圖 1 中的 ❷ )給 Alice。

Alice 依據下列步驟比對雙方的控制訊號和回傳比對結果：

(f). Alice 接收到 Bob 的控制訊號$B$後比對 Alice 的控制訊號$A$，產生比對結果$C = \{c_{2n-1}, c_{2n-2}, \ldots, c_1, c_0\}$, where $c_i = \begin{cases}1, a_i = b_i\\0, a_i \neq b_i\end{cases}$，並且通過經典通訊傳送比對結果$C$ (即圖 1 中的 ❸ )給 Bob。除此之外，Alice 可以根據比對結果$C$取得具有相同控制訊號(即$c_i = 1$)的量子位元值作為協商後的結果。

Bob 依據下列步驟根據比對結果取得協商後的結果：

(g). Bob 可以根據比對結果$C$取得具有相同控制訊號(即$c_i = 1$)的量子位元值作為協商後的結果。

*2) 實際案例*

為展示 BB84 協定，本研究以協商 4 個位元長度的訊息為例，所以總共需要用到 8 個量子位元，協商過程如表 I 所示。需注意的是，這僅是一個示意，實際應用在 AES-256 時，需要用到 512 個量子位元。

Alice 首先依據下列步驟產生和傳送量子訊號：

(a). Alice 隨機產生 8 個量子位元值，通過部分量子位元被操作 Pauli-X 閘後變換為$|1\rangle$，表示式為$Q_A = \{|1\rangle, |1\rangle, |0\rangle, |0\rangle, |1\rangle, |0\rangle, |0\rangle, |1\rangle\}$。

TABLE I.     BB84 協定實際案例
(符號說明:H 為 HADAMARD 閘、U 為均勻疊加態、Y/N 為標記是否一致)

| 步驟 | $q_7$ | $q_6$ | $q_5$ | $q_4$ | $q_3$ | $q_2$ | $q_1$ | $q_0$ |
|---|---|---|---|---|---|---|---|---|
| (a). Alice 隨機產生 $2 \times n$ 個量子位元值$Q_A$ | 1 | 1 | 0 | 0 | 1 | 0 | 0 | 1 |
| (b). Alice 隨機產生 $2 \times n$ 個控制訊號$A$ | H | I | H | I | H | I | H | I |
| (c). Alice 通過**量子通訊**(即圖 1 中的 ❶)傳送 $2 \times n$ 個量子位元$Q_T$給 Bob | $U^-$ | 1 | $U^+$ | 0 | $U^-$ | 0 | $U^+$ | 1 |
| (d). Bob 隨機產生 $2 \times n$ 個控制訊號$B$通過經典通訊傳送**控制訊號$B$**(即圖 1 中的 ❷ )給 Alice | H | H | I | I | H | H | I | I |
| (e). Bob **量測** $2 \times n$ 個量子位元$Q_B$和紀錄結果 | 1 | U | U | 0 | 1 | U | U | 1 |
| (f). Alice 標記每個控制訊號是否一致,通過經典通訊傳送**比對結果$C$** (即圖 1 中的 ❸ )給 Bob,同時 Alice 取得具有相同控制訊號的量子位元值 | 1 | 0 | 0 | 1 | 1 | 0 | 0 | 1 |
| (g). Bob 取得**具有相同控制訊號的量子位元值** | 1 |   |   | 0 | 1 |   |   | 1 |

(b). Alice 隨機產生 8 個控制訊號,表示式為 $A = \{H,I,H,I,H,I,H,I\}$,有部分量子位元會被操作 Hadamard 閘。其中,$H$ 表示有操作 Hadamard 閘,而 $I$ 表示不操作 Hadamard 閘。

(c). Alice 對 8 個量子位元值$Q_A$根據 8 個控制訊號$A$各別操作後產生欲傳送的量子訊號 $Q_T = \{|U^-\rangle,|1\rangle,|U^+\rangle,|0\rangle,|U^-\rangle,|0\rangle,|U^+\rangle,|1\rangle\}$,並且通過量子通訊傳送量子訊號$Q_T$(即圖 1 中的 ❶ )給 Bob。其中,$|U\rangle$表示為均勻疊加態。例如:第 7 個量子位元有操作 Hadamard 閘,則$H|q_7\rangle = |U^-\rangle$、第 6 個量子位元沒有操作 Hadamard 閘,則$I|q_6\rangle = |1\rangle$。

Bob 依據下列步驟量測量子訊號和傳送控制訊號:

(d). Bob 隨機產生 8 個控制訊號,表示式為 $B = \{H,H,I,I,H,H,I,I\}$。其中,$H$ 表示有操作 Hadamard 閘,而 $I$ 表示不操作 Hadamard 閘。例如:第 7 個量子位元 Bob 有操作 Hadamard 閘,而 Alice 有操作 Hadamard 閘,則$HH|q_7\rangle = |1\rangle$可以還原為$|1\rangle$、第 6 個量子位元 Bob 有操作 Hadamard 閘,但 Alice 沒有操作 Hadamard 閘,則$HI|q_6\rangle = |U\rangle$會得到均勻疊加態$|U\rangle$,無法肯定是$|0\rangle$或$|1\rangle$。

(e). Bob 對 8 個量子位元值$Q_T$根據 8 個控制訊號$B$各別操作後產生量測量子訊號 $Q_B = \{|1\rangle,|U\rangle,|U\rangle,|0\rangle,|1\rangle,|U\rangle,|U\rangle,|1\rangle\}$,並且通過經典通訊傳送控制訊號$B$ (即圖 1 中的 ❷ )給 Alice。

Alice 依據下列步驟比對雙方的控制訊號和回傳比對結果:

(f). Alice 接收到 Bob 的控制訊號$B$後比對 Alice 的控制訊號$A$,產生比對結果$C = \{1,0,0,1,1,0,0,1\}$,並且通過經典通訊傳送比對結果$C$ (即圖 1 中的 ❸ )給 Bob。除此之外,Alice 可以根據比對結果$C$取得具有相同控制訊號(即$c_i = 1$)的量子位元值作為協商後的結果,即$\{1,0,1,1\}$。

Bob 依據下列步驟根據比對結果取得協商後的結果:

(g). Bob 可以根據比對結果$C$取得具有相同控制訊號(即$c_i = 1$)的量子位元值作為協商後的結果$\{1,0,1,1\}$。

3) 安全性討論

目前已有部分文獻討論 BB84 協定協商過程中被監聽的安全性[20],所以本節主要討論兩種中間人監聽和嘗試為偽造量子訊號的可能性。

本節假設有中間人 Eve 可以監聽 Alice 和 Bob 之間的量子通訊和經典通訊訊號,並且具備能偽造量子通訊和經典通訊訊號的能力。其中,主要討論有中間人 Eve 是否能在監聽訊息 ❶ 後能偽造訊息 ⓐ 來達到攻擊的效果,如圖 2 所示。

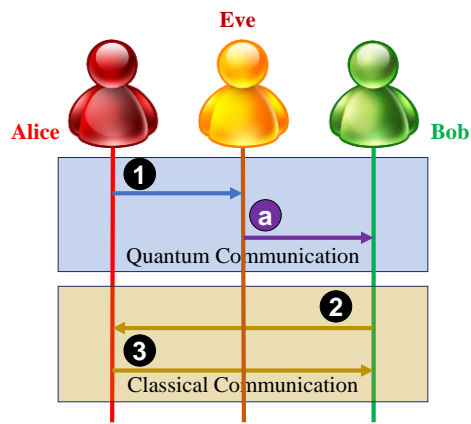

Fig. 2.  BB84 協定中間人監聽及其安全性

由於 Eve 並不知道 Alice 隨機產生$2 \times n$個量子位元值$Q_A$,也不知道 Alice 隨機產生$2 \times n$個控制訊號$A$,所以 Eve 無法預期哪幾個量子位元有被操作過 Hadamard 閘。因此,Eve 只能隨機猜測,並且對部分量子位元操作 Hadamard 閘,所以在本例中設備全部不操作 Hadamard 閘(如表 II 所示)和全部操作 Hadamard 閘(如表 III 所示)的情境下做討論。

在表 II 中,可以觀察到 Eve 可以正確得到 Alice 沒操作過 Hadamard 閘的那些量子位元值,但 Alice 有操作過

TABLE II. BB84 協定中間人監聽和嘗試偽造手法 1 (符號說明：H 為 Hadamard 閘、U 為均勻疊加態)

| 步驟 | $q_7$ | $q_6$ | $q_5$ | $q_4$ | $q_3$ | $q_2$ | $q_1$ | $q_0$ |
|---|---|---|---|---|---|---|---|---|
| (c). Alice 通過**量子通訊**(即圖 1 中的 ❶)傳送 **2×n 個量子位元**$Q_T$給 Bob | $U^-$ | 1 | $U^+$ | 0 | $U^-$ | 0 | $U^+$ | 1 |
| Eve 對全部量子位元不操作 Hadamard 閘 | I | I | I | I | I | I | I | I |
| Eve 量測 2×n 個量子位元和紀錄結果 | U | 1 | U | 0 | U | 0 | U | 1 |
| Eve 偽造和傳送 **2×n 個量子位元**$Q_T'$給 Bob(即圖 1 中的 ⓐ) | $U^-$ | 1 | $U^+$ | 0 | $U^+$ | 0 | $U^-$ | 1 |

TABLE III. BB84 協定中間人監聽和嘗試偽造手法 2 (符號說明：H 為 Hadamard 閘、U 為均勻疊加態)

| 步驟 | $q_7$ | $q_6$ | $q_5$ | $q_4$ | $q_3$ | $q_2$ | $q_1$ | $q_0$ |
|---|---|---|---|---|---|---|---|---|
| (c). Alice 通過**量子通訊**(即圖 1 中的 ❶)傳送 **2×n 個量子位元**$Q_T$給 Bob | $U^-$ | 1 | $U^+$ | 0 | $U^-$ | 0 | $U^+$ | 1 |
| Eve 對全部量子位元操作 Hadamard 閘 | H | H | H | H | H | H | H | H |
| Eve 量測 2×n 個量子位元和紀錄結果 | 1 | U | 0 | U | 1 | U | 0 | U |
| Eve 偽造和傳送 **2×n 個量子位元**$Q_T'$給 Bob(即圖 1 中的 ⓐ) | U | 1 | U | 1 | U | 0 | U | 0 |

Hadamard 閘的那些量子位元值卻被無法得知正確值。因此，當 Eve 要再偽造量子訊號時，Alice 有操作過 Hadamard 閘的那些量子位元有一半可能會被猜錯，所以 Eve 要偽造正確量子訊號的機率僅有$\frac{1}{2^n}$。例如：第 3 個量子位元原本是$|q_3\rangle = |U^-\rangle$，猜錯為$|U^+\rangle$、第 1 個量子位元原本是$|q_1\rangle = |U^+\rangle$，猜錯為$|U^-\rangle$。

同理，在表 III 中，可以觀察到 Eve 可以正確得到 Alice 有操作過 Hadamard 閘的那些量子位元值，但 Alice 沒操作過 Hadamard 閘的那些量子位元值卻被無法得知正確值。因此，當 Eve 要再偽造量子訊號時，Alice 沒操作過 Hadamard 閘的那些量子位元有一半可能會被猜錯，所以 Eve 要偽造正確量子訊號的機率同樣僅有$\frac{1}{2^n}$。例如：第 4 個量子位元原本是$|q_4\rangle = |0\rangle$，猜錯為$|1\rangle$、第 0 個量子位元原本是$|q_0\rangle = |1\rangle$，猜錯為$|0\rangle$。

由此可知，BB84 協定可以通過量子疊加態來建立安全性，並且協定過程中間人 Eve 偽造訊息後，Alice 和 Bob 可以察覺異樣，進而得知有被監聽。

### III. 本研究提出的量子密碼學中間人攻擊技巧

本研究發現在兩種攻擊情境將有機會攻破 BB84 協定，所以本節將針對這兩種攻擊情境做深入描述，並且各別討論其安全威脅。其中，本節中假設 Alice 為發送端、Bob 為接收端、以及 Eve 為中間人，並且 Eve 可以監聽 Alice 和 Bob 之間的量子通訊和經典通訊訊號，並且具備能偽造量子通訊和經典通訊訊號的能力。

*A. 中間人攻擊情境 1：發送端多次發送量子訊號*

在此攻擊情境中，主要由發送端發送兩次以上相同的量子訊號，將可能被中間人攻擊成功。

*1) 攻擊技巧*

假設 Alice 發送兩次量子訊號$Q_T$，並且 Eve 第一次監聽時對全部量子位元不操作 Hadamard 閘，以及 Eve 第二次監聽時對全部量子位元操作 Hadamard 閘，如圖 3 所示。

後續 Eve 可以對兩次監聽到的訊息做分析，各別去除均勻疊加態的量子位元值後將可能得到 Alice 隨機產生的$2 \times n$個量子位元值$Q_A$。

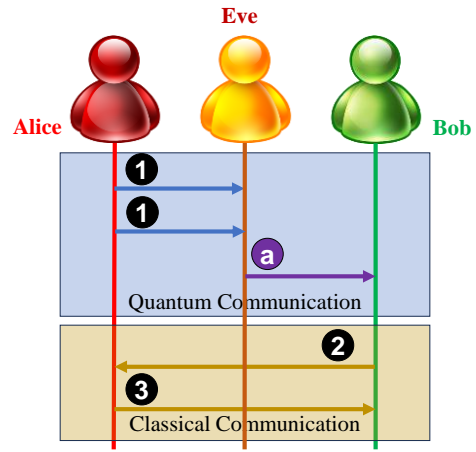

Fig. 3. 本研究提出的中間人攻擊情境 1

*2) 實際案例*

假設 Alice 發送兩次量子訊號 $Q_T = \{|U^-\rangle, |1\rangle, |U^+\rangle, |0\rangle, |U^-\rangle, |0\rangle, |U^+\rangle, |1\rangle\}$ (如表 I 所示)，並且 Eve 可監聽和量測這些量子訊號。

Eve 在第一次監聽時，對全部量子位元不操作 Hadamard 閘，可以得到$\{|U\rangle, |1\rangle, |U\rangle, |0\rangle, |U\rangle, |0\rangle, |U\rangle, |1\rangle\}$ (如表 II 所示)，從這個結果可以確定$\{q_6, q_4, q_2, q_0\} = \{|1\rangle, |0\rangle, |0\rangle, |1\rangle\}$。Eve 在第二次監聽時，對全部量子位元操作 Hadamard 閘，可以得到$\{|1\rangle, |U\rangle, |0\rangle, |U\rangle, |1\rangle, |U\rangle, |0\rangle, |U\rangle\}$ (如表 III 所示)，從這個結果可以確定$\{q_7, q_5, q_3, q_1\} = \{|1\rangle, |0\rangle, |1\rangle, |0\rangle\}$。

由於 Eve 在兩次量測後分別得到$\{q_6, q_4, q_2, q_0\} = \{|1\rangle, |0\rangle, |0\rangle, |1\rangle\}$和$\{q_7, q_5, q_3, q_1\} = \{|1\rangle, |0\rangle, |1\rangle, |0\rangle\}$，所以可以組合出 Alice 隨機產生的$2 \times n$個量子位元值$Q_A =$

TABLE IV. BB84 協定中間人攻擊情境 2 實際案例
(符號說明：H 為 HADAMARD 閘、U 為均勻疊加態、Y/N 為標記是否一致)

| 步驟 | $q_7$ | $q_6$ | $q_5$ | $q_4$ | $q_3$ | $q_2$ | $q_1$ | $q_0$ |
|---|---|---|---|---|---|---|---|---|
| (d). Bob 隨機產生 $2 \times n$ 個控制訊號$B$通過經典通訊傳送控制訊號$B$(即圖 4 中的 ❷)給 Alice | H | H | I | I | H | H | I | I |
| (a). Alice 隨機產生 $2 \times n$ 個量子位元值$Q_A$ | 1 | 1 | 0 | 0 | 1 | 0 | 0 | 1 |
| (b). Alice 隨機產生 $2 \times n$ 個控制訊號$A$ | H | I | H | I | H | I | H | I |
| (c). Alice 通過量子通訊(即圖 4 中的 ❶)傳送 $2 \times n$ 個量子位元$Q_T$給 Bob | $U^-$ | 1 | $U^+$ | 0 | $U^-$ | 0 | $U^+$ | 1 |
| Eve 根據監聽的控制訊號$B$建立對應的控制訊號 | H | H | I | I | H | H | I | I |
| Eve 量測 $2 \times n$ 個量子位元和紀錄結果 | 1 | U | U | 0 | 1 | U | U | 1 |
| Eve 偽造 $2 \times n$ 個量子位元$Q_T''$通過量子通訊(即圖 4 中的 ⓑ)傳送給 Bob | $U^-$ | 0 | $U^-$ | 0 | $U^-$ | 0 | $U^-$ | 1 |
| (e). Bob 量測 $2 \times n$ 個量子位元$Q_B$和紀錄結果 | 1 | U | U | 0 | 1 | U | U | 1 |
| (f). Alice 標記每個控制訊號是否一致，通過經典通訊傳送比對結果$C$(即圖 4 中的 ❸)給 Bob，同時 Alice 取得具有相同控制訊號的量子位元值 | 1 | 0 | 0 | 1 | 1 | 0 | 0 | 1 |
| (g). Bob 取得具有相同控制訊號的量子位元值 | 1 | | | 0 | 1 | | | 1 |

{|1⟩, |1⟩, |0⟩, |0⟩, |1⟩, |0⟩, |0⟩, |1⟩}。除此之外，通過比對第一次監聽結果，也可以得到 Alice 的控制訊號$A$。

*3) 安全威脅*

由於 Eve 已經完全破解出 Alice 隨機產生的$2 \times n$個量子位元值$Q_A$和 Alice 的控制訊號$A$，所以 Eve 可以偽裝成 Alice 來欺騙 Bob。Eve 將可以偽造出正確的量子訊號 $Q_T' = \{|U^-\rangle, |1\rangle, |U^+\rangle, |0\rangle, |U^-\rangle, |0\rangle, |U^+\rangle, |1\rangle\} = Q_T$，並且發送給 Bob，也就是圖 3 中的訊息 ⓐ 和訊息 ❶ 一致。並且，由於 Alice 和 Bob 無法偵測出錯誤，所以無法發現協商過程被監聽，Eve 攻擊成功。

*B. 中間人攻擊情境 2：接收端提早發送控制訊號*

在此攻擊情境中，主要由接收端提早發送其控制訊號，將可能被中間人攻擊成功。

*1) 攻擊技巧*

假設 Bob 提早發送其控制訊號$B$，並且 Eve 監聽經典通訊訊號時取得控制訊號$B$，如圖 4 所示。後續 Eve 可以根據控制訊號$B$偽造出可以欺騙 Bob 的量子訊號$Q_T''$。

當 Eve 監聽來自 Alice 傳送的量子訊號時，Eve 可以扮演 Bob 角色，運用與 Bob 相同的控制號來操作和量測結果。之後再根據量測後的結果，把已經可以得到明確值的部分根據控制訊號$B$再還原為原本的值，無法得到明確值的部分則隨機產生值後根據控制訊號$B$操作。再最後偽造的量子訊號$Q_T''$傳送給 Bob，雖然圖 4 中的訊息 ⓑ 和訊息 ❶ 不一致，但卻足以欺騙 Bob。

*2) 實際案例*

假設 Bob 發送控制訊號$B = \{H, H, I, I, H, H, I, I\}$，並且由 Eve 監聽經典通訊訊號時取得控制訊號$B$。當 Alice 發送量子訊號 $Q_T = \{|U^-\rangle, |1\rangle, |U^+\rangle, |0\rangle, |U^-\rangle, |0\rangle, |U^+\rangle, |1\rangle\}$ (如表 I 所示)，Eve 可根據來操作控制訊號$B$和量測這些

量子訊號，並且得到{|1⟩, |U⟩, |U⟩, |0⟩, |1⟩, |U⟩, |U⟩, |1⟩}。因此，Eve 可以得到預期要讓 Bob 量測到的結果(即$Q_B$)。

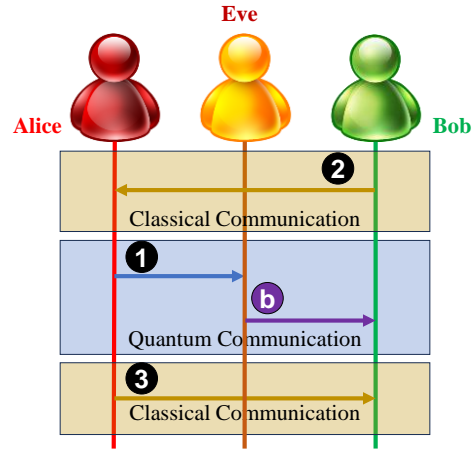

Fig. 4. 本研究提出的中間人攻擊情境 2

接下來，Eve 可以根據量測結果$Q_B$和控制訊號$B$來偽造量子訊號$Q_T''$傳送給 Bob。其中，每個量子位元值的操作說明如下，並且結果如表 IV 所示：

- 第 7 個量子位元的量測結果是|1⟩，並且 Bob 控制訊號為$H$，所以根據控制訊號操作後可得 $H|1\rangle = |U^-\rangle$。

- 第 6 個量子位元的量測結果是|U⟩，並且 Bob 控制訊號為$H$，所以表示原本是|0⟩或|1⟩其中一種，假設隨機設定為|0⟩。

- 第 5 個量子位元的量測結果是|U⟩，並且 Bob 控制訊號為$I$，所以表示原本是|$U^+$⟩或|$U^-$⟩其中一種，假設隨機設定為|$U^-$⟩。

- 第 4 個量子位元的量測結果是$|0\rangle$，並且 Bob 控制訊號為$I$，所以根據控制訊號操作後可得$I|0\rangle = |0\rangle$。
- 第 3 個量子位元的量測結果是$|1\rangle$，並且 Bob 控制訊號為$H$，所以根據控制訊號操作後可得$H|1\rangle = |U^-\rangle$。
- 第 2 個量子位元的量測結果是$|U\rangle$，並且 Bob 控制訊號為$H$，所以表示原本是$|0\rangle$或$|1\rangle$其中一種，假設隨機設定為$|0\rangle$。
- 第 1 個量子位元的量測結果是$|U\rangle$，並且 Bob 控制訊號為$I$，所以表示原本是$|U^+\rangle$或$|U^-\rangle$其中一種，假設隨機設定為$|U^-\rangle$。
- 第 0 個量子位元的量測結果是$|1\rangle$，並且 Bob 控制訊號為$I$，所以根據控制訊號操作後可得$I|1\rangle = |1\rangle$。

*3) 安全威脅*

在前一節中可以觀察到偽造的量子訊號$Q_T''$與 Alice 發送的量子訊號$Q_T$不一致(也就是圖 4 中的訊息 ⓑ 和訊息 ❶ 不一致)，但卻因為後續只取得具有相同控制訊號的量子位元值，所以偽造的量子訊號$Q_T''$足以欺騙 Bob。例如，在前述案例中(如表 IV 所示)，可以觀察到偽造的量子訊號$Q_T''$中的第 6 個、第 5 個、第 2 個量子位元值與 Alice 發送的量子訊號$Q_T$不一致，但協商後的值主要採用的第 7 個、第 4 個、第 3 個、第 0 個量子位元值。並且，由於 Alice 和 Bob 無法偵測出錯誤，所以無法發現協商過程被監聽，Eve 攻擊成功。

IV. 本研究提出的避免量子密碼學中間人攻擊的策略

有鑑於本研究提出的兩種中間人攻擊情境將可能對現行的量子密碼學造成安全威脅，所以本節將各別針對不同的攻擊技巧提出防禦策略。

*A. 中間人攻擊情境 1*

在此攻擊情境中，主要由發送端發送兩次以上相同的量子訊號，將可能被中間人攻擊成功。因此，為提高安全性，必須限制傳輸發送端發送同一個量子訊號時，只能發送一次。

雖然此策略可以避免中間人攻擊情境 1 的攻擊，然而在目前量子通訊和量測的硬體技術限制下可能存在誤差而導致協商失敗。因此，為了提升協商的成功率，必須提升硬體量測技術，以及可能需要更多的量子位元來做糾錯和修正使用。

*B. 中間人攻擊情境 2*

在此攻擊情境中，主要由接收端提早發送其控制訊號，將可能被中間人攻擊成功。因此，為提高安全性，必須限制接收端必須在做完量子訊號量測後再發送其控制訊號。

雖然此策略可以避免中間人攻擊情境 2 的攻擊，然而必須完全確認發送端不再發送量子訊號時再發送控制訊號給發送端，以避免被中間人攻擊成功。

V. 結論與未來研究

為建立安全的量子通訊，所以陸續有多個量子密碼學方法和量子金鑰分配協定被提出。然而，目前的量子金鑰分配協定(如：BB84 協定)仍可能存在安全漏洞。因此，本研究提出兩種中間人攻擊情境來破解 BB84 協定，並且提供實例論證如何破解 BB84 協定。此外，本研究亦針對這兩種中間人攻擊情境提出限禦策略，提供後續量子密碼學開發者參考。

本研究主要聚焦於討論 BB84 協定的安全性，但現行仍有不同的量子金鑰分配協定，未來可朝不同的量子金鑰分配協定進行討論。此外，BB84 協定主要用到量子疊加態來提供加密，而未來可以分析量子糾纏態的安全性，並且探索破解基於量子糾纏態密碼學的可能性。